\documentclass[lettersize,journal]{IEEEtran}
\usepackage{amsmath,amsfonts}
\usepackage{algorithmic}
\usepackage{cite}
\usepackage{array}
\usepackage[caption=false,font=footnotesize,labelfont=rm,textfont=rm]{subfig}
\usepackage{textcomp}
\usepackage{stfloats}
\usepackage{url}
\usepackage{verbatim}
\usepackage{graphicx}
\usepackage{xcolor}
\usepackage{multirow}
\usepackage{amsmath}
\usepackage{hyperref}
\def\BibTeX{{\rm B\kern-.05em{\sc i\kern-.025em b}\kern-.08em
		T\kern-.1667em\lower.7ex\hbox{E}\kern-.125emX}}
\usepackage{balance}
\begin{document}
	\title{\huge{Rotatable Antenna System Empowered Low-Altitude Economy: Opportunities and Challenges}}
\author{ 
	Shuaijun Li, Jie Tang,~\IEEEmembership{Senior Member,~IEEE}, Beixiong Zheng,~\IEEEmembership{Senior Member,~IEEE}, Lipeng Zhu,~\IEEEmembership{Member,~IEEE}, Cui Yang, Nan Zhao,~\IEEEmembership{Senior Member,~IEEE}, Xiu Yin Zhang,~\IEEEmembership{Fellow,~IEEE}, and Kai-Kit Wong,~\IEEEmembership{Fellow,~IEEE}} 
\maketitle 
	
\begin{abstract}
Low-altitude economy (LAE) is an emerging technological paradigm that enables continuous airspace coverage at multiple altitudes by providing highly reliable data connectivity for numerous low-altitude applications. However, existing networks cannot sufficiently support LAE development, as current base stations (BSs) are primarily designed for terrestrial users and lack the capability to provide continuous coverage at low altitudes. To overcome these challenges, rotatable antenna system (RAS) is introduced in LAE, enabling flexible beamforming by dynamically adjusting the boresight of directional antennas to extend low-altitude coverage and enhance the stability of data transmission. In this article, we first provide an overview of RAS-empowered LAE applications, including low-altitude communication, sensing, control, and computation. Then, we present two practical RAS deployment strategies for LAE scenarios, namely RAS-aided multi-BS and multi-unmanned aerial vehicle (UAV) cooperative coverages, as well as provide detailed discussions on their system architectures and performance benefits. Additionally, key design issues of RAS in LAE are discussed, including channel modeling and estimation, cellular access and interference cancellation, as well as RAS configuration and boresight optimization. Finally, we demonstrate the performance gains of RAS in LAE networks through experimental and simulation results.
	
\end{abstract}
	
%	\begin{IEEEkeywords}
%		
%	\end{IEEEkeywords}

\section{Introduction}
With great advancements in next-generation wireless communication technologies and the rapid expansion of the low-altitude aircraft industry, the low-altitude economy (LAE) is experiencing unprecedented growth opportunities. LAE encompasses economic activities in airspace below 1000 meters (m), driven by both manned and unmanned aerial vehicles (UAVs). It is characterized by an extensive industrial chain, wide-ranging applications, strong growth potential, and significant economic impact, with various applications in aviation, logistics, and smart cities\cite{Jiang2025Non-Terres}. Globally, many countries and regions are actively exploring and developing LAE. To effectively manage and optimize the use of low-altitude airspace resources, governments and regulatory bodies have introduced policies and regulations to support and oversee industry development, thereby creating new opportunities and challenges for society and the economy.

Traditional wireless networks primarily provide ground coverage, whereas LAE network requires three-dimensional (3D) aerial coverage. Moreover, LAE network needs to incorporate active sensing capabilities to detect real-time environmental changes and ensure seamless monitoring of target areas in 3D space. Recent technological advancements have drawn increasing attention to low-altitude integrated sensing and communication (ISAC) due to its superior performance and flexible deployment\cite{Tang2025Co-ISAC}. Furthermore, low-altitude ISAC places higher demands on device miniaturization and operational flexibility. By actively sensing the surrounding environment, it facilitates optimized UAV trajectory planning and flexible resource allocation. However, in air-to-ground scenarios, strong line-of-sight (LoS) propagation, while ensuring high communication quality, inevitably introduces co-channel interference and reduces the effective  channel rank\cite{Zeng2019UAV}, thereby degrading detection accuracy and diminishing spatial multiplexing gain.

Meanwhile, existing mobile networks are primarily designed to serve terrestrial users, and traditional down-tilted base stations (BSs) are insufficient for aerial coverage. In addition, the fixed orientation of antennas at the BS further hinders the tracking capability of low-altitude moving targets\cite{Wang2025Non-Terre-model}. Although antenna sidelobes can offer limited signal coverage in low-altitude areas, ground obstacles often cause LoS blockage, resulting in signal instability, reduced coverage, and beam misalignment. To meet the service demands of different altitude scenarios, current low-altitude networks adopt a hierarchical coverage strategy. In airspace above 100 m, a low-altitude private network can be established by deploying dedicated aerial BSs, which can be installed on elevated structures such as towers, mountaintops, or rooftops, or mounted on platforms such as UAVs and airships to provide wide coverage and flexible deployment\cite{Jiang2025Non-Terres}. However, such deployments inevitably results in increased costs. In the airspace below 100 m, although the existing  terrestrial network  can be reused, it still fails to meet the flexible coverage demands of low-altitude areas. Furthermore, satellite communication and mobile ad hoc networks can provide partial support for the development of LAE. However, the long distance between satellites and low-altitude UAVs or ground BSs results in significant signal attenuation and transmission delay, while the bulky size and high cost of satellite equipment further limit their deployment on UAVs. By contrast, mobile ad hoc networks rely more heavily on multi-hop relaying, which increases energy consumption, reduces spectral efficiency, and substantially increases end-to-end latency.

On the other hand, the diversity of low-altitude application scenarios, the dynamic of the airspace environment, and the high mobility of aerial devices present significant challenges in building an efficient and stable LAE network\cite{Zhao2025GenerativeAW}. Additionally, the rapid development of LAE, along with the sharp increase in the number of aerial vehicles, presents a bottleneck for LAE networks in supporting massive connectivity and ensuring ultra-reliable, low-latency communication. Although multiple-input multiple-output (MIMO) technology can enhance system performance by increasing the number of antennas, its high cost and deployment complexity significantly constrain its application in miniaturized devices for low-altitude scenarios. Furthermore, current MIMO technologies primarily rely on fixed antennas with boresights and positions that remain unchanged after deployment, thereby lacking the capability to adapt to spatial channel dynamics\cite{Kit2024FAS}. 

\begin{figure*}[!t]
%	\vspace{-0.8cm}
	\centering
	\includegraphics[width=1.0\textwidth]{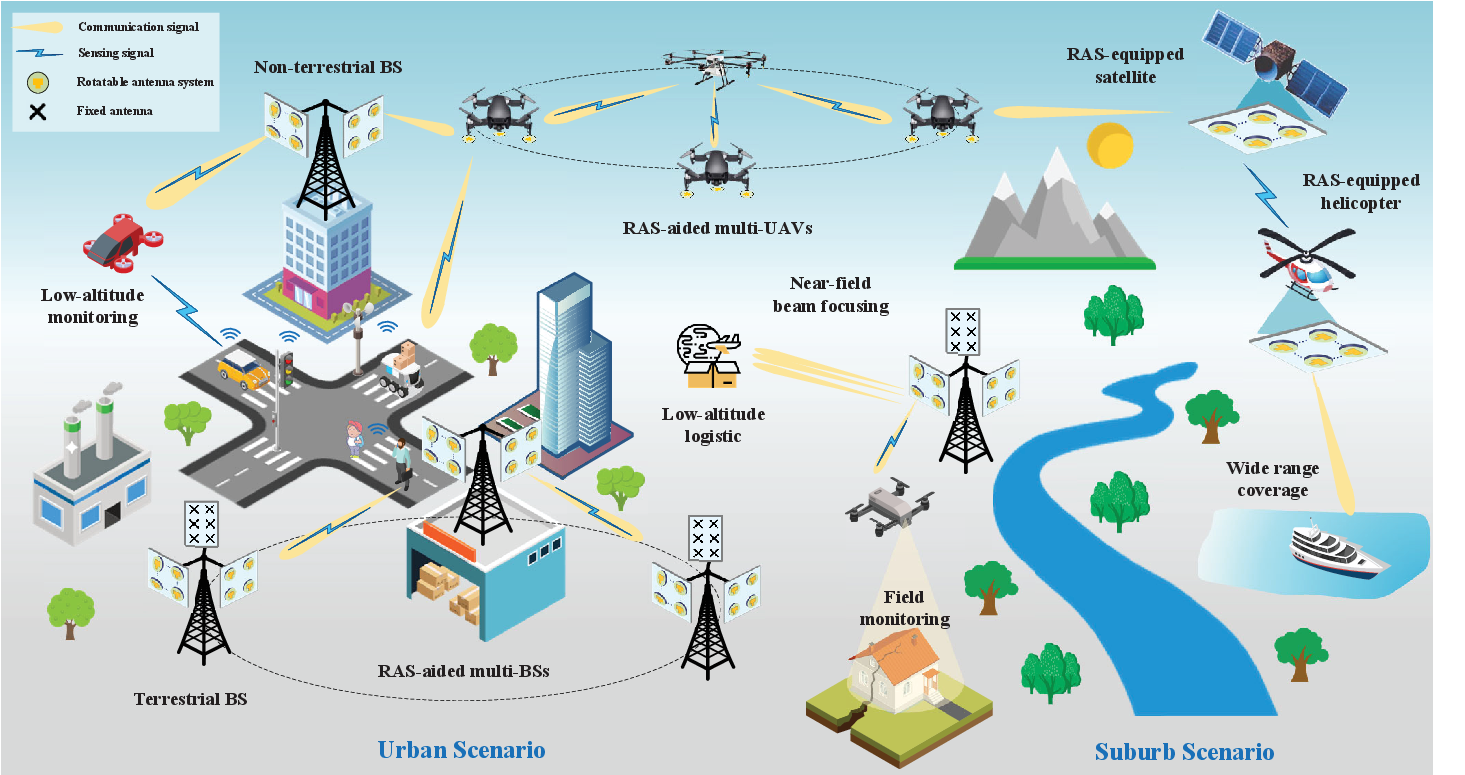}
	\caption{Illustration of RAS applications in LAE.}
	\label{Scenario}
\end{figure*} 

Recently, to overcome the limitations of fixed-antenna architecture in spatial degrees of freedom (DoFs) and channel adaptability, flexible antenna technology with dynamically adjustable positions and orientations has attracted increasing attention. By dynamically reconfiguring the positions and/or shapes of antenna elements, fluid antenna systems (FAS)\cite{Kit2024FAS} and movable antenna (MA)\cite{Zhu2025MA} can exploit spatial diversity within a predefined space to create favorable channel conditions without increasing the number of antennas. This introduces additional DoFs and actively reshapes the wireless channel to enhance communication and/or sensing performance. Consequently, they significantly reduce the hardware complexity and energy consumption. However, the potential of using FAS/MA in LAE networks is not well understood, which motivates this article. 
 
Motivated by the above, this article proposes an innovative rotatable antenna system (RAS)  within the FAS framework, aiming to enhance the performance and coverage of LAE networks in a cost-effective manner. Unlike existing flexible antenna technologies, which typically  require additional space, RAS only involves local rotational adjustment while the antenna’s position remains fixed \cite{Zheng2025RA_magzine}. This can be easily implemented using compact mechanical or integrated electronic designs, ensuring compatibility with existing wireless systems and efficient deployment in space- and cost-limited scenarios. 
In further detail, RAS is not a new concept in antenna technology. Mechanically controlled RAS can be traced back to radar systems of the mid-20th century, where they primarily involved the rotation of entire antenna arrays and were mostly used in radar and navigation applications. In contrast, the considered RAS supports not only the independent rotation of individual antennas but also the collective rotation of the entire array, with each mode exhibiting distinct response characteristics.

Generally speaking, from the perspective of LAE, the hierarchical coverage strategy integrating terrestrial and non-terrestrial networks enables a certain degree of aerial coverage but also results in substantial deployment costs. In addition, due to the limited spatial efficiency of fixed antennas, the coverage and tracking of low-altitude mobile targets experience certain challenges\cite{Xu2025LAE}. In this regard, RAS can be co-deployed with existing BSs and/or aerial nodes to flexibly steer the radiation pattern toward desired directions. As shown in Fig. \ref{Scenario},, the LAE has a wide range of applications in logistics, agriculture, and tourism, with core demands covering communication, sensing, control, and computing. These functions are no longer regarded as independent modules but constitute an integral component of a unified wireless infrastructure. Although this integration improves system efficiency, it inevitably entails performance trade-offs resulting from the shared use of spectral, spatial, and computational resources.  Notably, the real-time reconfiguration capability of RAS provides new opportunities to balance performance across these domains. Besides, RAS-aided multi-BS and multi-UAV cooperative coverages are two pivotal strategies in RAS-empowered LAE network.  The main differences and competitive advantages of RAS-based LAE, compared to fixed-antenna based LAE, are summarized in Table \ref{table_comp}. The key functions of RAS-empowered LAE are elaborated as follows:

    \begin{itemize}
       	\item \textbf{Low-altitude communication:} RAS can flexibly adjust the boresight of each antenna according to the spatial distribution of low-altitude users/targets, enabling dynamic allocation of antenna resources across diverse application scenarios. This not only improves channel conditions and enhances spatial multiplexing gains, but also mitigates interference and supports flexible beamforming in dynamic low-altitude environments. Furthermore, by exploiting more spatial DoFs with fewer antennas, more stable link quality and higher throughput can be achieved, while communication quality feedback can further guide the selection of sensing beams.
       	
     	\item \textbf{Low-altitude sensing:} RAS can dynamically steer its boresight to direct beams toward different low-altitude service areas, achieving wide angular coverage and high-precision real-time environmental awareness. Furthermore, the radiation patterns of all RASs can be simultaneously focused on a common spatial target, forming a spotlight-like beam suitable for near-field sensing. RAS can also perform multi-angle observation and multi-dimensional scanning, effectively reducing blind spots and capturing multiple target features. This enhances sensing coverage and resolution, enables efficient multi-target sensing with fewer antennas, and allows the collected sensing data to support Low-altitude channel state prediction and beam tracking.
     	
    	\item \textbf{Low-altitude control and computation :} RAS provides an efficient spatial task decoupling mechanism, enabling different antenna resources to independently adjust their orientations while concurrently supporting low-altitude environmental sensing, data transmission, control, and computation across multiple spatial directions. In addition, by optimizing the quality and directionality of transmission links from distributed devices to edge servers and steering beams toward computing nodes, the system can reduce transmission delays and energy consumption during task offloading, thereby supporting efficient computing resource scheduling and reliable data transmission.	
    	
    \end{itemize}
    
\renewcommand{\arraystretch}{1.5} % default is 1.0
\renewcommand{\arraystretch}{1.5} % default is 1.0
\begin{table*}[!t]
%	\vspace{-0.3cm}
	\caption{Comparison of RAS-Based LAE and Fixed-Antenna Based LAE  \label{tab:comp}}
	%	\vspace{-0.1cm}
	\centering
	\begin{tabular}{|m{2.0cm}<{\centering}|m{2.3cm}<{\centering}|m{2.2cm}<{\centering}|m{2.5cm}<{\centering}|m{2.8cm}<{\centering}||m{2.9cm}<{\centering}|} 	
		\hline
		\bf Deployment Strategy &  \bf Hardware Architecture &  \bf Beamforming Implementation & \bf Communication Comparison  &  \bf Sensing Comparison  &  \bf Control and Computation Comparison \\ 
		\hline \hline
		RAS-based LAE  & Antenna/array-level rotation, mechanically/electronically driven & Antenna/array-level beamforming & High interference mitigation, spatial multiplexing; and flexible beamforming & High sensing resolution/coverage; and multi-target/dimensional sensing & High transmission/computation offloading rate; and offloading link reliability \\
		\hline
	Fixed-antenna based LAE & Fixed sector antenna  & Array-level beamforming & Limited spatial DoFs, beamforming, and interference suppression & Limited coverage and sensing capabilities &  Higher antenna overhead for same performance  \\
		\hline
	\end{tabular}
%	\vspace{-0.2cm}
	\label{table_comp}
\end{table*}

\section{RAS-Aided Multi-BS Cooperative Coverage}
    In LAE scenarios, continuous coverage of low-altitude areas and real-time tracking of moving targets are fundamental requirements for enabling low-altitude communication, sensing, control, and computation. However, traditional fixed-antenna BSs provide limited spatial coverage and result in blind zones, while a single BS lacks the capability for multi-target sensing and multi-dimensional feature extraction. To address this issue, RAS can be co-deployed with existing BSs and dedicated aerial BSs to complement coverage, thereby mitigating blind spots and enhancing robust sensing capabilities in dynamic airspace environments.
    
    As shown in Fig.\ref{Scenario}, RAS-aided terrestrial and non-terrestrial BSs provide cooperative coverage, enabling diverse applications in LAE scenarios and extending service from ground-facing coverage to support for terrestrial and aerial users/targets. In this case, RAS can steer the entire array pattern toward the desired direction to achieve array-level beamforming, while also reconfiguring the radiation pattern of each antenna for element-level beamforming, thereby enhancing the reliability of multi-target detection and suppressing interference. Moreover, the low-altitude environment is typically characterized by a limited number of scatterers and dominant LoS paths, especially under near-field conditions, which can reduce channel rank and degrade spatial multiplexing gains. By dynamically adjusting the boresight, RAS arrays increase the angular separation between signal paths, thereby improving spatial resolution and enabling multi-stream spatial multiplexing. Furthermore, to better identify and track low-altitude moving targets, RAS array can steer the boresight of all antennas toward the same target, thereby significantly enhancing sensing resolution and localization accuracy. On the other hand, RAS-aided multi-BS scan the target area from different directions through independent antenna rotation. Sharing and integrating sensing data from multiple BSs not only reduces resource waste caused by redundant detection and excessive searching but also significantly enhances system coverage and observation angle diversity.
    
\begin{figure*}[!t]
   		\vspace{-0.5cm}
   	\centering
   	\includegraphics[width=1.0\textwidth]{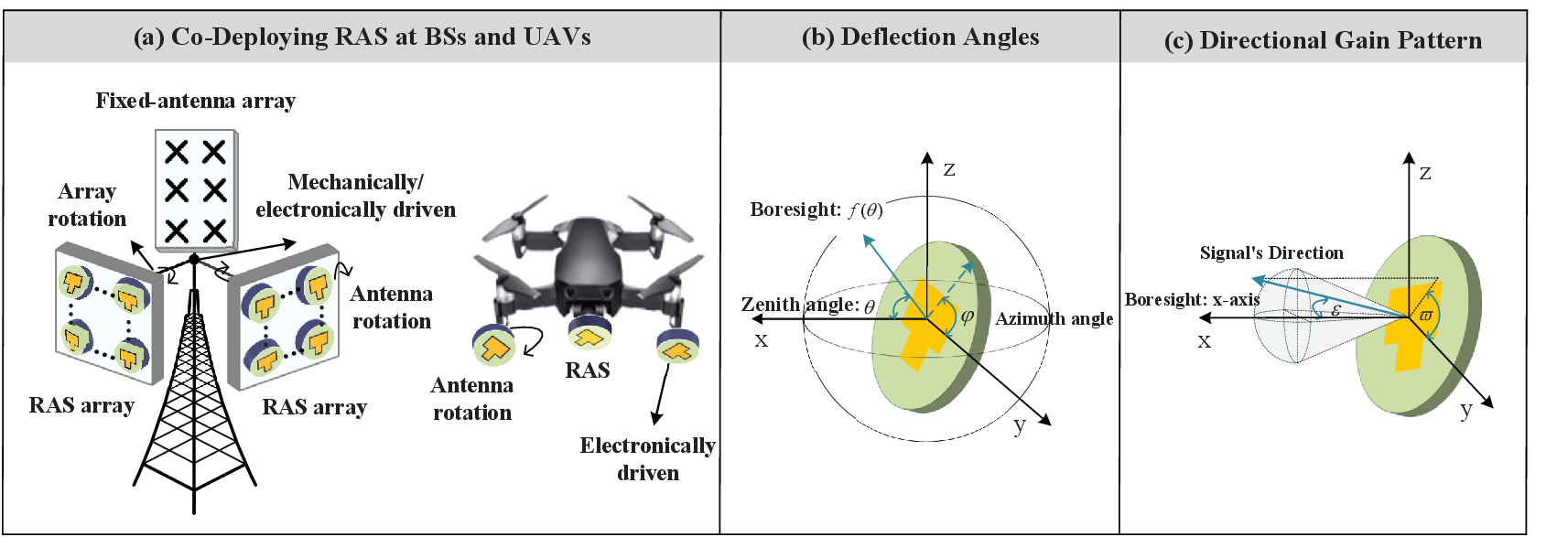}
   	\caption{Architectures and model for RAS.}
   	\label{Architecture}
   \end{figure*}

    Although RAS-aided multi-BS cooperative coverage shows significant potential in low-altitude scenarios, further discussion of system design details is necessary to enable practical implementation. Firstly, in terms of the hardware architecture, as shown in Fig. \ref{Architecture}a, RAS can be integrated into existing BSs and UAVs to enhance compatibility with current networks. In this setup, fixed sector antennas continue to serve terrestrial users, while the co-deployed RAS extends coverage to support terrestrial and aerial users/targets. Specifically, to reduce hardware overhead and design complexity, a hybrid architecture combining array-level and antenna-level rotation is adopted, thereby enhancing rotational flexibility and coverage. However, array-level rotation inherently alters the antenna radiation pattern and may require compensation, such as rotation calibration, to maintain the desired system performance. To balance the broader boresight adjustment range provided by mechanical driven with the shorter response time of electronic driven, a co-designed RAS architecture that integrates mechanisms can be employed. Among them, mechanically driven approaches rotate the antenna using actuators such as servo or micro-electromechanical systems (MEMS) motors, offering a wide range but at the cost of latency, mechanical wear, and milliwatt-level power consumption, with response times from microseconds to milliseconds. In contrast, electronically driven RAS keep the antenna fixed and steer the boresight through tunable materials (e.g., liquid crystals, PIN diodes) \cite{Zheng2025RA_magzine}, enabling faster millisecond-scale response and solid-state implementation but within a narrower range. Although mechanical methods incur slightly higher latency, both approaches remain practical depending on the real-time demands of LAE applications.

    Next, we present the general system model of RAS, providing an intuitive illustration of the antenna's boresight adjustment\cite{Zheng2025RA}. As shown in Fig. \ref{Architecture}b, the zenith angle $\theta$ denotes the deviation of the RAS's boresight direction. To avoid interference among neighboring RASs, $\theta$ should be constrained in $[0,\pi/2]$. The azimuth angle $\varphi$ is defined as the angle between the projection of the RAS's boresight direction onto the $y$-$z$ plane and the $y$-axis. To facilitate the characterization of the RAS's 3D boresight direction, the normalized boresight direction vector is denoted as $f(\theta)$. In addition, Fig. \ref{Architecture}c shows the radiation pattern gain in different directions, where the effective gain of the RAS is jointly determined by the radiation pattern and the signal's arrival/departure angle. Specifically, the general directional gain of each RAS in the signal deflection direction can be expressed as
    $G_0cos(\varepsilon)$, where $G_0$ denotes the maximum gain in the boresight direction, and  $(\varepsilon,\varpi)$ represents a pair of incident angle of any spatial point relative to the current boresight of the RAS.
     
    Last but not least, achieving precise target tracking and efficient beam scanning scheduling also plays a pivotal role. Generally, when channel state information (CSI) is known, the array’s beamforming weights and each antenna’s boresight directions can be jointly optimized to point toward the target position, thereby enhancing sensing resolution/coverage. In the absence of reliable prior information, a coarse-to-fine scanning strategy can be adopted to achieve a trade-off between scanning resolution and latency. Moreover, RAS-aided collaborative dynamic beam coverage by multiple BSs is expected to significantly enhance system performance, while also increasing the complexity of coordination schemes. In particular, efficiently and reliably exchanging and integrating sensing data among multiple BSs to achieve high-quality seamless coverage remains an open problem that requires further research.
    
\section{RAS-Aided Multi-UAV Cooperative Coverage}
Air-to-ground coverage is vital for the stable operation of collaborative services in LAE, as it directly affects the efficiency of communication, sensing, air traffic management, and ground coordination. However, since UAVs, as small-payload devices, cannot accommodate large-scale antennas for flexible beam coverage and a single UAV inevitably suffers from limitations in sensing range and communication rate, these challenges must be carefully addressed in system design. Besides, in the LAE, control and nonpayload communication (CNPC) is designed to ensure the safety, reliability, and efficiency of flight operations, while payload communication focuses on the timely transmit and/or receive mission-related data. The two differ in communication methods and spectrum requirements. RAS provides a lightweight and flexible solution that not only improves the coordination performance of multi-UAV systems but also balances data transmission rate and quality under different communication demands, without relying on large-scale antenna arrays.

As shown in Fig. \ref{Scenario}, UAVs equipped with RASs leverage their high mobility and flexibility, along with the RAS’s capability to fully exploit 3D space, to perform 360-degree scanning of the surrounding environment. Specifically, the RAS mounted on the UAV adjusts its boresight direction to perform multi-dimensional sensing and dynamically scan the surrounding environment, with the goals of optimizing the UAV trajectory, enhancing air-to-ground coverage, and mitigating interference caused by strong LoS paths\cite{Mu2023UAV}. Moreover, RAS enables flexible far-field beam steering and near-field beam focusing, facilitating adaptive identification and tracking of high-mobility targets while maintaining high-quality links with ground and aerial nodes. This capability allows the UAV swarm to more accurately sense the topological environment, estimate relative distances and angles between neighboring UAVs, avoid intra-swarm collisions, and establish a robust swarm topology. Furthermore, as low-altitude ISAC frequencies extend to the millimeter-wave and terahertz bands and antenna array sizes increase, near-field effects become increasingly pronounced. Near-field beams can focus energy at a specific angle and distance while providing additional resolution in the distance domain. Therefore,  investigating RAS performance in near-field communications is crucial for mitigating inter-user interference, improving spectral efficiency, and supporting emerging paradigms such as position-division multiple access.

RAS-aided multi-UAV cooperative coverage shows strong potential in LAE, but its system design requires further discussion to support practical implementation. Given the small payload and high mobility of UAVs, dynamic beam control and low-overhead scanning within the swarm are essential to ensure quality of service. As shown in Fig. \ref{Architecture}a, RAS can be integrated into UAVs primarily as antenna elements, while maintaining a small aperture and a compact electronically driven design. By flexibly rotating the radiation pattern, RAS enables element-level beamforming and scans multiple spatial directions sequentially or concurrently with fewer antennas, thereby enhancing moving target detection and interference suppression\cite{Xu2025LAE}. Moreover, given the limited computational capacity of UAVs, RAS can dynamically scan regions of interest through rotation to construct environmental knowledge maps and select appropriate codebooks based on environmental changes, enabling efficient beam switching\cite{Zhu2025MA}. 

In the LAE, cellular-connected UAVs may not only cause severe interference to multiple co-channel BSs but also experience strong interference from non-associated BSs. Their high mobility frequently causes handovers and leads to time-varying wireless backhaul links with ground BSs or users. Nevertheless, the flexible movement of UAVs in 3D airspace introduces additional DoFs in trajectory design, thereby offering new opportunities for performance optimization in low-altitude networks.
Additionally, the RAS can perform multi-dimensional sensing of the surrounding environment through flexible beam control and scanning scheduling strategies, enabling the extraction of target features such as direction, velocity, size, and orientation. Furthermore, considering the limited onboard computational resources and stringent real-time constraints of UAVs, offloading computation-intensive tasks to edge servers reduces both computational and transmission burdens on the swarm, thereby enhancing overall response efficiency\cite{Meng2024UAV-Enabled}.

Finally, the LAE network is susceptible to factors such as weather changes, air traffic, and unexpected events, making it essential to ensure quality of service (QoS) and robustness in dynamic environments. In particular, low-altitude networks inevitably suffer from co-channel interference and multipath fading, as interfering signals typically arrive from multiple directions. RAS can exploit its boresight direction reconfigurability to rotate the main lobe, mitigating interference more directly and efficiently while maintaining high gain toward the target user. However, in practical applications, RAS may require exhaustive searching across all possible discrete levels. Additionally, multi-view estimation or sensing in dynamic environments further increases the complexity of space-time signal processing. To address these challenges, learning-based frameworks, such as neural networks and reinforcement learning, can be employed to enhance signal processing capabilities under complex conditions, thereby improving the overall stability of the LAE network\cite{Zhao2025GenerativeAW}.
		
\section{Design Issues}
Although RAS offers significant benefits and promising potential, several issues remain in implementing RAS-empowered LAE networks to enhance coverage, communication, and sensing performance. This section will focus on key issues in RAS-empowered LAE network, including channel modeling and estimation, cellular access and interference cancellation, and RAS configuration and boresight optimization. It will also address the main challenges currently encountered and propose forward-looking research directions and solutions.

\subsection{Channel Modeling and Estimation}
In low-altitude scenarios, the wireless channel is highly complex and dynamic, making accurate CSI essential for effective boresight control and for unlocking the full potential of RAS in LAE. Among them, channel modeling characterizes the statistical properties of the channel to support system design and performance analysis, while channel estimation enables real-time tracking of channel states to adapt to dynamic variations. Moreover, the rotation of the RAS changes the antenna radiation pattern and polarization, thereby affecting the channel gain in path transmission. Specifically, it alters the impinging angle of the electromagnetic wave, which in turn modifies the radiation pattern, while polarization changes can be represented by the polarization response matrix in the channel model, further influencing the channel gain. On the other hand, RAS rotation introduces spatial correlation and hardware coupling effects, which are mainly caused by circuit coupling between adjacent elements, leading to inaccuracies in channel modeling. This issue can be mitigated by leveraging circuit and antenna theory together with nonlinear models, including matching network design and isolation techniques, to achieve a more accurate characterization of mutual coupling.
Fortunately, although RAS dynamically adapts to channel variations by rotating its radiation pattern, the antenna position remains fixed during rotation, thereby preserving the stable physical topology formed by the transceivers and the propagation environment. Consequently, RAS only needs to estimate a limited set of environmental features to support channel estimation or sensing. At the same time, conventional channel estimation techniques designed for fixed antennas remain applicable, ensuring compatibility with existing wireless infrastructure.

In addition, existing 3GPP channel standards primarily focus on communication propagation and lack the capability to accurately characterize the integrated channel features of low-altitude communication and sensing\cite{3GPP2024}. Thus, the 3GPP framework can be extended to integrate RAS, sensing targets, UAV terminals, and environmental objects into a unified propagation environment, enabling the joint generation of background and target channels for RAS-empowered LAE. In contrast, channel estimation, which relies on actual received signals, compensates for the limitations of channel models in practical scenarios. As the number of RAS antennas and candidate rotation angles increases, channel estimation becomes more complex and costly, while the channel power distribution remains uneven. By exploiting the directional sparsity of RAS channels or applying data-driven artificial intelligence (AI) algorithms, CSI can be estimated with more efficiency and accuracy\cite{Zhao2025GenerativeAW}.
	
\subsection{Cellular Access and Interference Cancellation}
In low-altitude scenarios, the high mobility of aircraft can cause frequent cell handovers and overlapping coverage. At the same time, strong LoS paths result in severe co-channel interference, which significantly affects the dynamic allocation of physical resource blocks (PRBs) across different cells\cite{kkwong2023access}. Unlike traditional systems with fixed antennas that rely primarily on filtering, precoding, and other signal processing or resource allocation techniques to suppress or cancel interference, RAS can dynamically adjust its boresight direction based on the location of interference sources, thereby mitigating inter-cell co-channel interference and multipath fading. To better support the dynamic allocation of PRBs across different cells, RAS must consider the diverse requirements of various tasks in terms of scan resolution, response latency, and angular accuracy during the scanning process. It should also balance the overhead and delay tolerance of scanning scheduling to ensure efficient PRB allocation and overall system performance.

In addition, a cell association mechanism and a multi-objective resource allocation strategy tailored to low-altitude scenarios are proposed. By jointly optimizing array beamforming weights, antenna boresight directions, and UAV trajectories, the LAE network effectively mitigates inter-cell interference, enhances coverage and sensing capabilities, and reduces both coverage overlap and handover frequency. Furthermore, a path optimization model is developed by integrating wireless network performance with flight trajectory characteristics, aiming to minimize path loss, reduce handover frequency, and stabilize the sensing signal-to-interferenceplus-noise ratio (SINR), thereby reducing coverage overlap and the occurrence of cellular handovers\cite{Xu2025LAE}. Building upon this, the resource allocation strategy for LAE networks is further explored by integrating ground communication and aerial user sensing needs, with the goal of improving resource reuse efficiency between ground and low-altitude users.

\subsection{RAS Configuration and Boresight Optimization}
In RAS-empowered LAE network, different application scenarios have specific requirements for RAS configurations. Depending on the specific tasks, it is necessary to flexibly select structural designs and deployment strategies, such as electronic or mechanical control methods, array-level or antenna-level rotational scales, and centralized or distributed deployments. Specifically, in wireless systems that prioritize spatial resolution, coverage range, and diversity of observation angles, distributed RAS deployment combined with a large array aperture is typically employed to improve sensing accuracy and coverage, while mitigating coupling effects during rotation and reducing adverse shadowing. However, in size-constrained applications with stringent requirements for compact array design and control methods, such as small payload devices like UAVs, closer
antenna spacing and electronically driven control are preferred to facilitate practical integration.
Although continuous tuning of RAS surface rotation provides maximum flexibility, practical mechanical systems such as rotary motors typically permit only discrete rotational adjustments. Continuous adjustment not only increases hardware cost, power consumption, and movement time but also makes finding the optimal solution computationally expensive due to the numerous discrete rotation combinations.

RAS can dynamically adjust its boresight direction based on the spatial distribution of users/targets, thereby enhancing directional gain and suppressing interference. By jointly optimizing the array beamforming weights and antenna boresight directions, RAS flexibly reconfigures the overall array pattern and/or the radiation pattern of individual antennas to achieve far-field directional control or near-field power concentration, thereby enabling accurate identification and tracking of high-mobility targets\cite{Zheng2025RA}. However, boresight direction optimization requires accurate CSI, which is difficult to obtain in practice. Thus, it is necessary to have a trade-off between optimality and computational complexity. Specifically, with perfect CSI, methods such as gradient descent, successive convex approximation (SCA), and alternating optimization (AO) can jointly optimize all RAS boresight directions. In contrast, under imperfect CSI, exhaustive search is typically employed to identify the optimal direction, although it often incurs significant time and energy overhead. Moreover, AI- and swarm intelligence–based optimization methods can be utilized to iteratively adjust the RAS's boresight direction, enabling more precise control.

\begin{figure*}[!t] 
	\centering
	\includegraphics[width=1.0\textwidth]{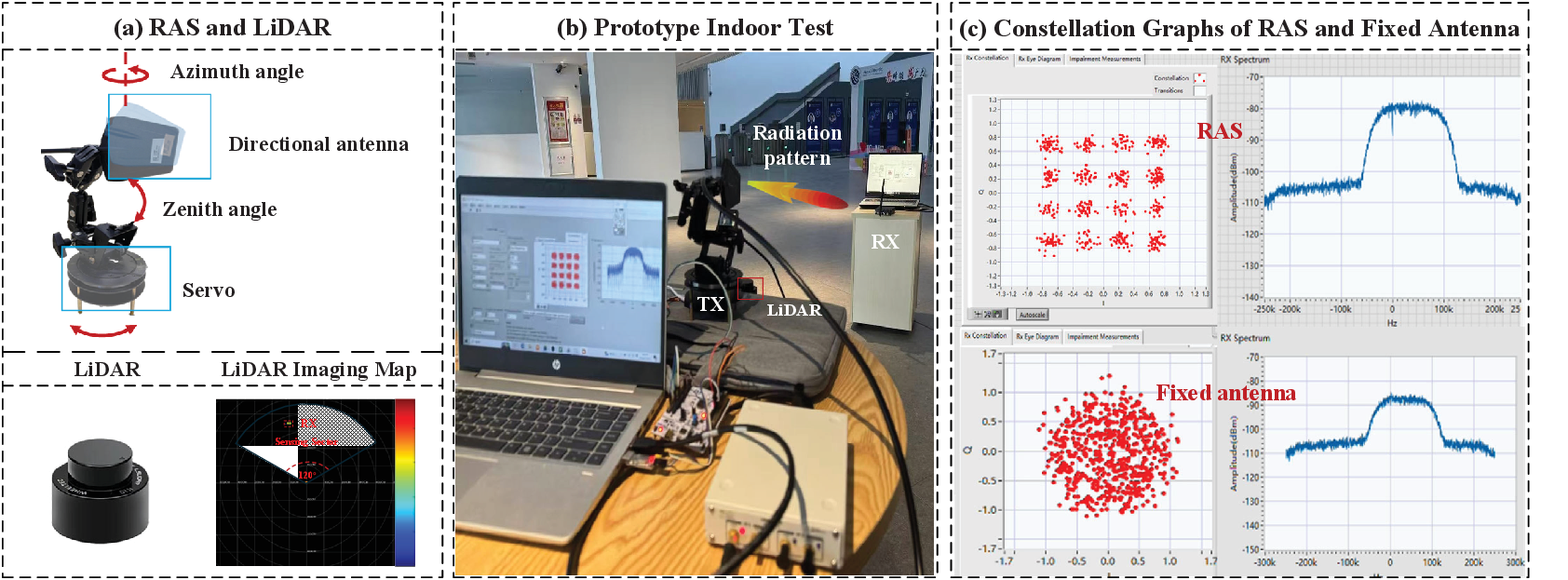} 
	\caption{ RAS prototype and experimental results.}
	\label{Prototype}
\end{figure*}

\section{Case Studies}
To demonstrate the superior performance of RAS in LAE networks, practical experiments and numerical simulations are conducted on RAS-aided communication and/or sensing. First, a mechanically driven RAS prototype is developed, and its performance gain over fixed antennas is evaluated through practical tests. Subsequently, simulations are performed to further validate the performance improvements of RAS in low-altitude scenarios.

\subsection{Experimental Results}
As shown in Fig. \ref{Prototype}, a prototype system was established in an indoor environment, comprising a universal software radio peripheral (USRP) for signal processing, a personal computer
(PC) for control and analysis, and a LiDAR to locate the RX, while an STM32 microcontroller adjusts the servo motor to align the RAS’s boresight with the RX based on its angle of arrival (AoA) information\cite{Zheng2025RA_demo}. The system operates at a carrier frequency of 5.8 GHz with 16-quadrature amplitude modulation (QAM) and a bandwidth of 100 kHz. It transmits at 10 dBm, with a noise power of approximately –95 dBm, over a distance of 4 m. Compared with a fixed-antenna system, the RAS produces a clearer, more regularly distributed constellation diagram, demonstrating significant improvement in received signal quality. Additionally, the RAS achieves a higher received signal power than the fixed-antenna
counterpart, with an approximate gain of 7 dBm. The experimental results above confirm the RAS’s capability to enhance directional gain and suppress interference.
 
\subsection{Numerical Results}
We consider a multi-user low-altitude ISAC system, where the BS employs a uniform planar array (UPA), with the antenna radiation pattern modeled according to\cite{Zheng2025RA}. Users and sensing targets are uniformly distributed in a 3D spherical annulus centered at the BS, with radial distances ranging from 30 m to 150 m. Within this setup, the system operates at a carrier frequency of 2.4 GHz, corresponding to a wavelength of $\lambda=0.125$ m, an average noise power of –80 dBm, and the minimum antenna spacing on each RAS is $\lambda/2$. To further demonstrate the advantages of RAS-empowered LAE, the proposed RAS, optimized based on statistical CSI, is compared with the baseline schemes.

\begin{figure}[!t]
	\centering
	\includegraphics[width=3.2in]{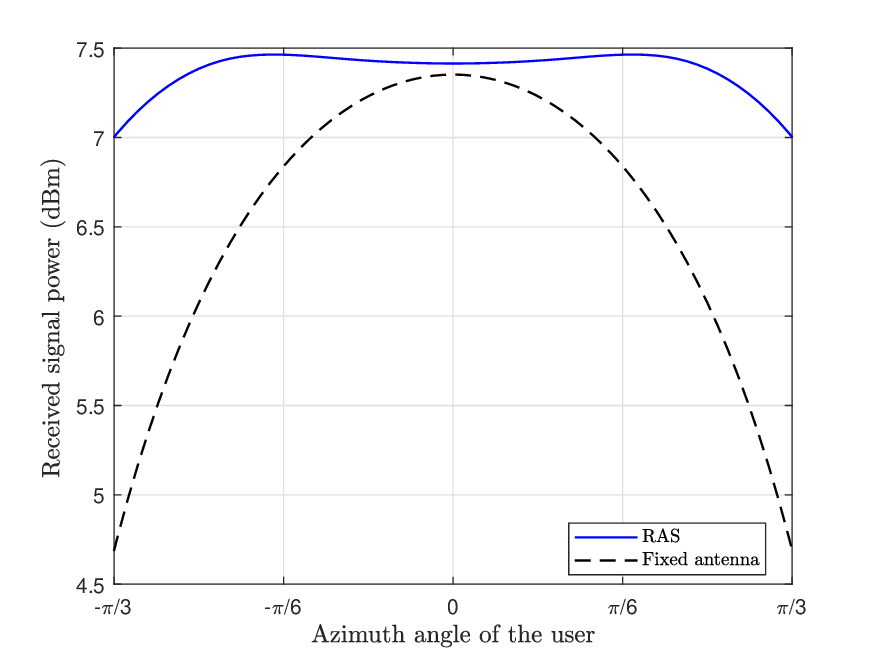}
	\caption{Comparison of received signal power between RAS and fixed antenna.}
	\label{Power}
%	\vspace{-0.35cm}
\end{figure}
\begin{figure}[!t]
	\centering
	\includegraphics[width=3.2in]{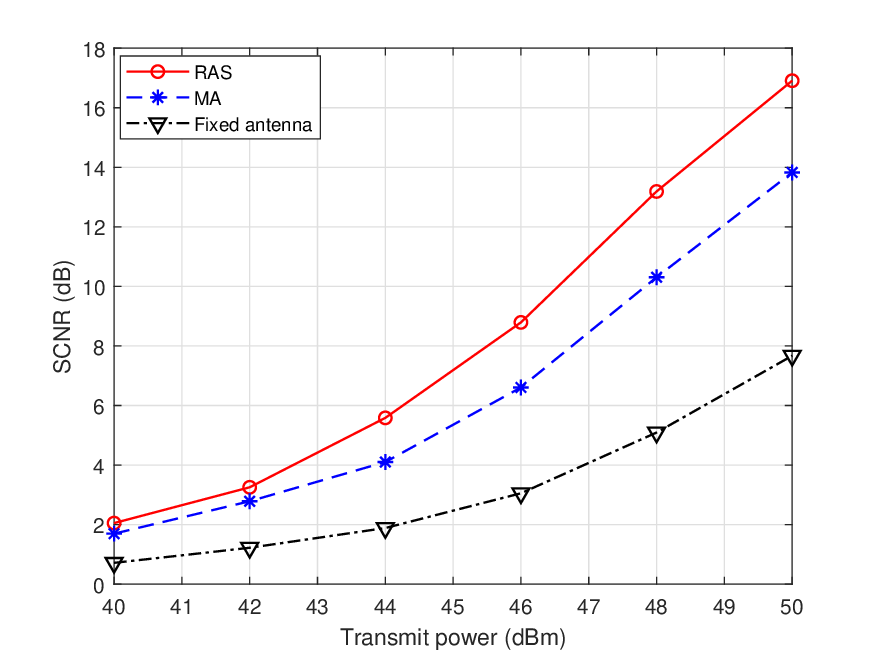}
	\caption{SCNR versus transmit power at the BS under different schemes.}
	\label{SCNR}
%	\vspace{-0.26cm}
\end{figure}

Fig. \ref{Power} depicts the received signal power of different schemes versus the azimuth angle. When the user’s zenith angle is fixed at $\pi/2$, the received power of the fixed-antenna system decreases significantly as $\varphi$ increases from 0 to $\pi/3$ or decreases to $-\pi/3$. This occurs because its array gain pattern is fixed, with radiated power concentrated only in the forward region. In contrast, RAS can dynamically steer the main lobe according to the user’s direction, thereby enhancing received power, suppressing interference, and maintaining stable connectivity in dynamic environments, which enables the BS to achieve more uniform and reliable reception across the entire angular range.

Fig. \ref{SCNR} compares the signal-to-clutter-and-noise ratio (SCNR) of different schemes versus the BS transmit power in the RAS-empowered low-altitude ISAC system. It is observed that the RAS  achieves a higher SCNR than the fixed-antenna and MA schemes under the same transmit power. This is attributed to the RAS’s greater spatial DOFs, which allows better adaptation to user distribution, increased antenna gain in the desired direction, and improved angular resolution and sensing accuracy. Moreover, as the transmit power increases, the performance gap between systems widens, as the SCNR becomes decreasingly limited by interference at higher transmission power or received signal-to-noise ratio (SNR).

\section{Conclusions}
This article provided a comprehensive overview of RAS-empowered LAE within the FAS framework, highlighting its potential to enhance airspace coverage and overall performance in low-altitude environments. We first introduced the applications of RAS in collaborative coverage involving multiple BSs and UAVs, outlining its fundamental principles and highlighting its advantages over conventional techniques. We then reviewed the key design issues in applying RAS to LAE networks, and proposed promising solutions to improve system performance and practical feasibility. Finally, case studies were provided to demonstrate substantial performance gains of RAS-empowered LAE networks. It is foreseeable that RAS-empowered LAE will open up new research avenues for achieving wide coverage and massive connectivity in future low-altitude scenarios.

\bibliographystyle{IEEEtran}

\bibliography{LAE_reference}

%%%%%%%%%%%%%%%%%%%%%%%%%%%%%%%%%%%%%%%%%%%%%%%%%%%%%%%%%%%%%%%%%

  \end{document}